\documentclass[runningheads]{llncs}

\usepackage{color}
\usepackage{graphicx} 
\usepackage{hhline} 
\usepackage{bbding} 
\usepackage{tabularx} 
\usepackage{ragged2e}
\usepackage{cite}
\usepackage{paralist} 
\usepackage{enumerate}
\usepackage{amsmath}

\usepackage[dvipsnames]{xcolor}
\usepackage{colortbl}
\usepackage{pgfplotstable}
\usepackage{multirow}
\usepackage{multicol}
\usepackage{booktabs}
\usepackage{array}
\usepackage{makecell}
\usepackage[export]{adjustbox}

\usepackage{pgfplots}
\usepackage{subcaption}
\pgfplotsset{compat=1.3} 

\newcommand{\orcid}[1]{\href{https://orcid.org/#1}{\textcolor[HTML]{A6CE39}{\aiOrcid}}}

\begin{document}
\title{Privacy-Preserving Detection of IoT Devices Connected Behind a NAT in a Smart Home Setup}

\titlerunning{Privacy-Preserving Detection of IoT Devices Connected Behind a NAT}

\author{Yair Meidan \inst{1}\orcidID{0000-0003-4865-2334} \and
Vinay Sachidananda\inst{2}\orcidID{0000-0001-9582-1538} \and
Yuval Elovici\inst{1}\orcidID{0000-0002-9641-128X} \and
Asaf Shabtai\inst{1}\orcidID{0000-0003-0630-4059}
}


\authorrunning{Meidan et al.}

\institute{Ben-Gurion University of the Negev, Beer-Sheva, Israel \email{\{yairme\}@post.bgu.ac.il, \{elovici,shabtaia\}@bgu.ac.il} \\
\and
iTrust - Singapore University of Technology and Design, Singapore \\ \email{sachidananda@sutd.edu.sg}}

%
%
%
\maketitle              

\begin{abstract}
Today, telecommunication service providers (telcos) are exposed to cyber-attacks executed by compromised IoT devices connected to their customers' networks. 
Such attacks might have severe effects not only on the target of attacks but also on the telcos themselves. 
To mitigate those risks we propose a machine learning based method that can detect devices of specific vulnerable IoT models connected behind a domestic NAT, thereby identifying home networks that pose a risk to the telco's infrastructure and availability of services. 
As part of the effort to preserve the domestic customers' privacy, our method relies on NetFlow data solely, refraining from inspecting the payload.
To promote future research in this domain we share our novel dataset, collected in our lab from numerous and various commercial IoT devices.

\end{abstract}



%

\section{Introduction}\label{sec:introduction}
The ability to launch massive distributed denial of service (DDoS) attacks via a botnet of compromised devices 
is an exponentially growing risk in the Internet of Things (IoT)~\cite{CNA2016DDoSIMDA,Rayome2017DDoSIoT}.
Such massive attacks, possibly emerging from IoT devices in home networks~\cite{Kambourakis2017TheArmies}, hit not only the target of the attacks, but also the infrastructure of telecommunication service providers (telcos) along the attack path as well. 
By virtue of the huge bandwidth that is assigned to customers nowadays
, the combined traffic surge from infected IoT devices that might hit the telco's infrastructure could eventually overload it. 
This might cause episodic downtime and serious backlashes in the form of widespread customer dissatisfaction.

Typically, IoT-based DDoS attacks rely on exploiting vulnerabilities of specific models of IoT devices~\cite{Kambourakis2017TheArmies}. 
In such cases most domestic customers who connect IoT devices to their home networks don't have the knowledge or means to handle ongoing attacks, and the burden of preventing them falls on the telco. 
To effectively scale and defend against IoT-based attacks launched from customers' premises, telcos can continuously monitor the traffic of their customers. 
Based on the monitored traffic, telcos can detect exploitation attempts, infections, and executed attacks on third parties, and then block these activities. 
However, this 
approach might be too late and result in service malfunctions and also hurt the telco's reputation. 
Due to these reasons, we propose a method for detecting connected vulnerable IoT models 
\textit{before they are compromised}.
Thus, in the case of DDoS attacks, our method can facilitate offloading of huge traffic amounts generated by an abundance of infected domestic IoT devices. 
In turn, this can prevent the combined traffic surge from hitting the telco's infrastructure, reduce the likelihood of service disruption and ensure continued service availability.
 
In this paper we propose a novel method for telcos to mitigate the above IoT-related risks posed by their domestic customers. 
It relies on monitoring the traffic of each smart home 
separately in order to verify: \textit{Is an IoT model, known to be vulnerable to a DDoS attack, connected to this network or not?} This method relies on NetFlow records and it does not violate the privacy of customers since it does not analyze traffic payloads.
A telco using our proposed method can (1) detect vulnerable IoT devices connected behind a NAT, and (2) use this information to take actions. 
We empirically evaluate our method on genuine NetFlow records collected in our lab for a period of ten days from numerous commercial smart home IoT devices.
We also compare our NetFlow-based method to two existing deNATing methods: (1) a domain-based method~\cite{guo2018ip} and (2) a method which is based on DNS IP-ID~\cite{orevi2018dns}; we evaluate them empirically on packet-level data collected simultaneously from the same network. 
Unlike some past studies which applied their methods to partially, questionably, or completely unlabeled datasets, our datasets are explicitly labeled with the device model. We share all of our datasets with the scientific community to promote future reproducible research, so given the ground truth labeling, both our study and future studies can be truthfully evaluated in terms of classification performance.

\section{Background}\label{sec:background}

\subsection{NATing, deNATing and IoT Identification Behind a NAT}\label{subsec:NATing_and_denating}

In home networks it is common~\cite{Gokcen2014CanFlows,Maier2011NATNetworks} to use NAT-enabled Wi-Fi routers. 
As part of \textit{NATing} the outbound traffic, the NAT routers effectively 'hide' the internal IP addresses of individual connected devices by replacing them with the router's external IP address.
Once NATed, it becomes difficult to correlate each packet to its packet stream from the outside.
As described by~\cite{orevi2018dns}, \emph{deNATing} is the reverse of \emph{NATing}, and it aims at re-identifying the communication flowing through a NAT. 
In Section~\ref{sec:related}, we survey existing deNATing methods and illustrate their shortcomings with regard to our use case.

In the IoT identification literature it is typically required to analyze chronological sequences of packets~\cite{Miettinen2017IoTIoT, Lopez-Martin2017NetworkThings} or sessions~\cite{meidan2017profiliot} for device (type) fingerprinting or even for the basic calculation of inter-arrival times~\cite{Mahalle2013a, radhakrishnan2015gtid}. 
When the traffic is NATed separating multiple packets into distinct timely sequences becomes a real challenge, thus the validity of existing IoT identification methods is undermined. 
To overcome this, we propose to use the popular NetFlow protocol (discussed next), which already aggregates packets internally into \textit{(Net)Flows}. 
Nonetheless, although NetFlow practically performs deNATing internally, the detection of specific IoT models based on a single NetFlow remains a definitive challenge.

\subsection{NetFlow as a Basis for Privacy-Preserving Traffic Analysis}\label{subsec:Netflow_traffic_analysis} 
In the domain of network traffic analysis several levels of data granularity are typically used to define an entity, such as a packet, a transaction, a session, a flow, a conversation window, etc.~\cite{callado2009survey, Bekerman2015UnknownClassification}. Among them, packet-level traffic analysis is a common approach, wherein deep packet inspection (DPI) is an accepted conduct~\cite{el2017survey}. 
This approach requires the capturing and analysis of highly-detailed network traffic data including the payload of each packet.
Although potentially informative for IoT model detection, there are some known disadvantages~\cite{li2013survey} to using DPI in terms of efficiency and privacy preserving, as follows. 

\begin{itemize}
    \item \textbf{Efficiency}:
    The collection and analysis of the entire raw traffic (including the payload) in networks of high traffic rates is technically challenging. For a telco, the traffic volume can reach up to multiple Gigabits per second and it is far from trivial to capture and analyze such tremendous amounts of data.
    \item \textbf{Privacy preserving}:
    DPI allows the telco access to its customers' personal information. Even more dangerously, in case this data is leaked, there is a privacy risk to the communicating parties, as this information might expose private data (e.g., video captures that are transmitted over HTTP). 
\end{itemize}

To address the above disadvantages we propose to use NetFlow~\cite{netflow2017} instead. NetFlow was integrated as a feature in Cisco's routers and provides the capability to summarize IP network traffic passing through an interface. 
By relying only on traffic statistics and metadata aggregated by NetFlow, we preserve the privacy of the communicating parties. We do not access the actual payload and we do not use this information at all in our method. 
In addition, collecting and analyzing only NetFlow's statistical aggregations instead of the raw data requires significantly less computation and storage, thus making the analysis more efficient. It is also worth mentioning that NetFlow is a common solution natively supported by most routers~\cite{verde2014no}, and it is acknowledged as the de-facto standard for compact representation of large traffic rates. Actually, NetFlow was also used in the past for security-related tasks like Botnets detection and deNATing and has already been introduced as a privacy preserving solution~\cite{verde2014no,abt2012towards,guo2018ip}.

\subsection{System and Threat Model}\label{subsec:threat_model} 

Our system model is a typical network setup found in smart homes. The IoT devices are connected to a gateway router providing an interface for connecting IP-enabled devices to the Internet. We assume that during the initial setup when the IoT devices connect to the network they possibly have security vulnerabilities that are not yet exploited. In our threat model, we assume that (1) our network setup is likely to be targeted with attacks (e.g., DDoS) which could be carried out by botnets such as Mirai; (2) The telco would like to observe the traffic emerging from the customers' premises which are NATed; (3) The telco is a passive listener who wishes to block the IoT devices which are vulnerable and susceptible to cyberattacks; and (4) The telco does not know which applications are running and will preserve privacy while monitoring the traffic.


\section{Research Goals and Contributions}\label{sec:research_goal_contribution}

We focus on the following questions: 
\begin{enumerate}
    \item Can we detect IoT models connected behind a NAT by analyzing NetFlows?
    \item Can we perform this detection without compromising the users' privacy?
    \item Can we provide satisfactory detection performance for a large variety of IoT models and validate our approach to be better than state of the art?
\end{enumerate}

We summarize our contributions as follows:
\begin{enumerate}
    \item To the best of our knowledge, we are the first to apply machine learning techniques to NATed network traffic for IoT model detection.
    \item The current state of the art relies on data sources and features that necessitate the analysis of high-definition data and thus compromise users' privacy.
    In our approach we preserve privacy, mostly by using only common meta-features extracted by NetFlow.
    \item We evaluate our method using genuine traffic data collected from various IoT devices, and demonstrate that we can detect IoT models behind a NAT. We also share both our NetFlow and pcap datasets with the scientific community.
\end{enumerate}

\section{Related Work}
\label{sec:related}

\subsection{Scope and Orientation}\label{subsec:scope_and_orientation}

Several prior NAT-related studies (for example,~\cite{Gokcen2014CanFlows}) 
 focused on identifying the presence of a NAT device in a network. 
We refer to them as non-deNATing, since they don't perform classical deNATing, i.e., they don't divide NATed traffic (all with the same source IP of the NAT router) into distinct packet streams for further analysis. 
As can be seen in Table~\ref{tab:past_motivation}, most of the other studies aimed at uncovering the identity and/or the quantity of the devices connected \textit{behind} such NAT devices, or the people who use them. 
Some motivations for doing so:
\begin{itemize}
    \item \textbf{Security}: Detecting attackers~\cite{Ori2008IdentifyingTechniques} or devices that are vulnerable~\cite{Patton2014UninvitedIoT, guo2018ip} or infected by a malware~\cite{orevi2018dns}.
    \item \textbf{Traffic management}: Applying policies such as parental browsing control or per device communication limits~\cite{orevi2018dns}, as well as 
    traffic interception.
    \item \textbf{Commerce}: Traffic profiling for targeted advertising~\cite{orevi2018dns}.
    \item \textbf{Privacy violation}: Inferring user connectivity~\cite{verde2014no} and behavior~\cite{Apthorpe2017}.
\end{itemize}

Our motivation is also security-oriented; it reflects the viewpoint of a telco wishing to defend against IoT botnets that might severely impact the communication availability. 
 Unlike prior studies whose subject of interest was (NATed) operating systems~\cite{orevi2018dns} or people and their behavior~\cite{Apthorpe2017,verde2014no,Ori2008IdentifyingTechniques,Savage2000PracticalTraceback}, our method is designed to detect connected IoT device models. 
Additionally, in contrast to some NAT-related studies performed in the past which addressed large-scale environments like smart cities~\cite{verde2014no,Patton2014UninvitedIoT,Savage2000PracticalTraceback,guo2018ip} and smart manufacturing~\cite{Ercolani2016ShodanVisualized,Gokcen2014CanFlows,Patton2014UninvitedIoT,guo2018ip}, we tailored our method to smart homes.
We evaluated it empirically using a wide range of popular home (consumer) IoT devices like smart light bulbs, sockets, and webcams, as well as laptops and smartphones.

\begin{table}
  \caption{The scope and orientation of previously-conducted deNAT-related studies}
  \label{tab:past_motivation}
  \includegraphics[width=\linewidth]{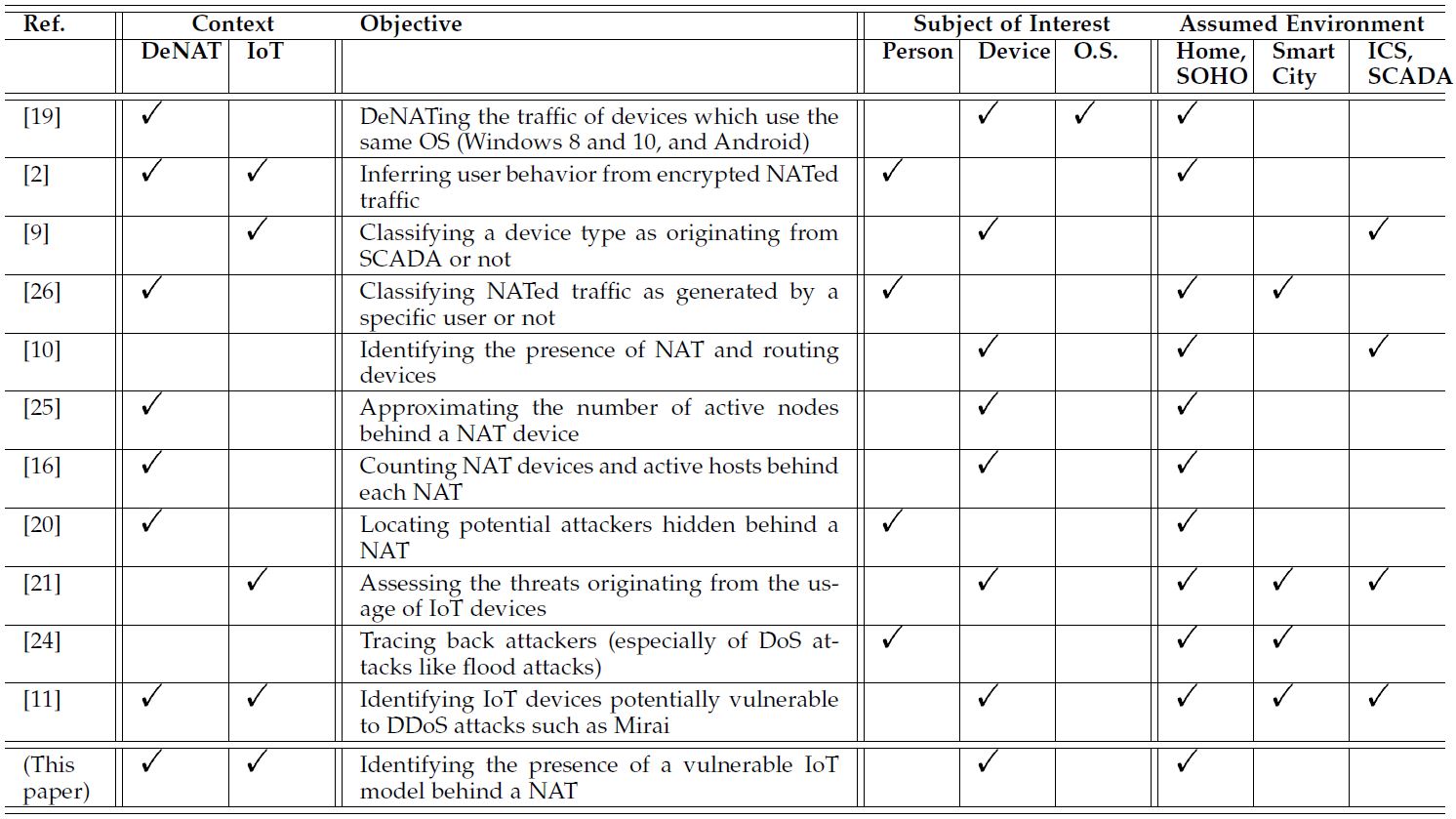}
\end{table}

\subsection{Methods and Evaluation}\label{subsec:methods_and_evaluation}

As summarized in Table~\ref{tab:past_techniques}, a variety of data sources and related methods have been proposed for deNATing.
Most of them rely on features extracted directly from the TCP/IP model, however in some cases they might fail to address our use case.
For instance, features from the application layer like the HTTP user agent~\cite{Gokcen2014CanFlows}, MSN transaction ID~\cite{Ori2008IdentifyingTechniques}, and DNS domains~\cite{Apthorpe2017} might not be available to the telco when the traffic is encrypted.
In contrast, we use NetFlow which extracts data from different layers (such as the network and transport layers which are usually not encrypted), as well as metadata like traffic rates and volumes. The TCP timestamp is another commonly used feature in deNATing~\cite{orevi2018dns,Tekeoglu2011ApproximatingDevice}, however its analysis requires a minimal number of packets, and it might even be disabled. 
Moreover, many IoTs use the UDP transport protocol, so the robustness of this feature becomes questionable.
In contrast, our NetFlow-based method can handle both TCP and UDP, so it covers a wider range of IoT models. 
Other approaches relied on the server(s) the IoT devices communicate with~\cite{Apthorpe2017} and the related DNS domains~\cite{guo2018ip}. 
With NetFlow, we can analyze all kinds of traffic and not just specific types such as DNS. 
Some works relied on the IP TTL~\cite{Maier2011NATNetworks} based on the assumption that a NAT device decrements its value by one, however the value at which it decrements might differ among various NAT devices. 
Others proposed using the open ports~\cite{Ercolani2016ShodanVisualized} to deduce the traffic's origin. 
However, these studies aimed at roughly distinguishing between SCADA and non-SCADA traffic. Our method is better suited for the given use case, as it proposes fine-grained classification. We have empirically shown that our method is effective at distinguishing between multiple IoT models, even among specific models of the same make (e.g., different D-Link webcams).

\begin{table}
  \caption{The methods used by previous deNAT-related studies}
  \label{tab:past_techniques}
  \includegraphics[width=\linewidth]{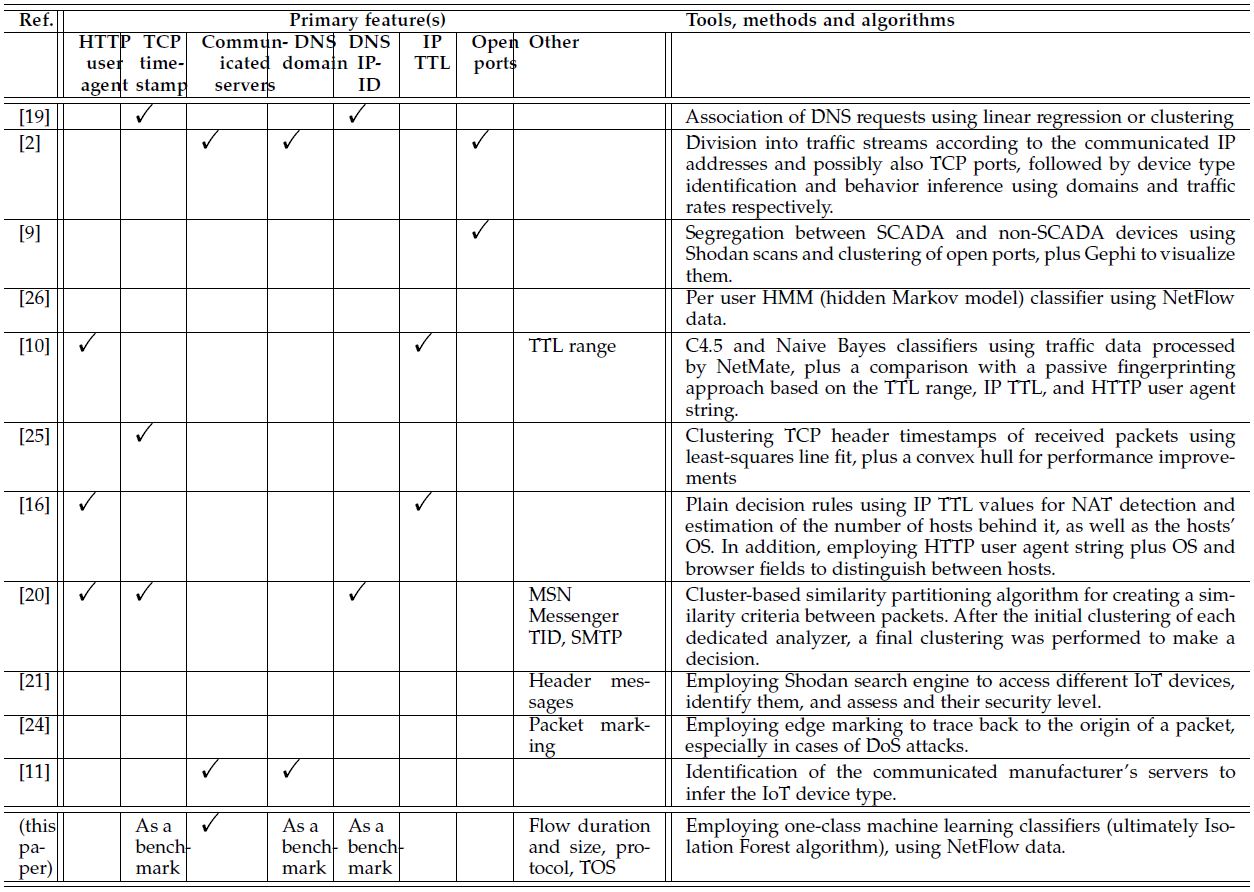}
\end{table}

For empirical evaluation, most prior studies used tcpdump log files, Shodan scans~\cite{Ercolani2016ShodanVisualized, Patton2014UninvitedIoT}, or simulated data~\cite{Savage2000PracticalTraceback}. 
In some cases, the dataset is not available for research reproduction~\cite{Gokcen2014CanFlows}; in some cases, it is questionably labeled~\cite{guo2018ip} or not at all~\cite{Maier2011NATNetworks}; and in most cases, it does not represent smart homes. Moreover, even if labeling is present, the data only comes from non-IoT hosts~\cite{Ori2008IdentifyingTechniques}, and the class does not reflect the device model~\cite{orevi2018dns}. 
In contrast, in our research we use NetFlow records, collected in our controlled home-like lab for a period of ten days from various IoT and non-IoT devices.
For benchmarking we also simultaneously collected pcap files, and most importantly, we explicitly labeled our data with the ground truth regarding the device models.
We believe that the novelty, authenticity, diversity, scope, and reliability of our publicly available datasets will facilitate future research and may serve as a benchmark for deNATing algorithms, specifically in the context of smart homes.




\section{Proposed Method}\label{sec:proposed_method}

\begin{figure}[ht]
\centering
\includegraphics[width=\textwidth]{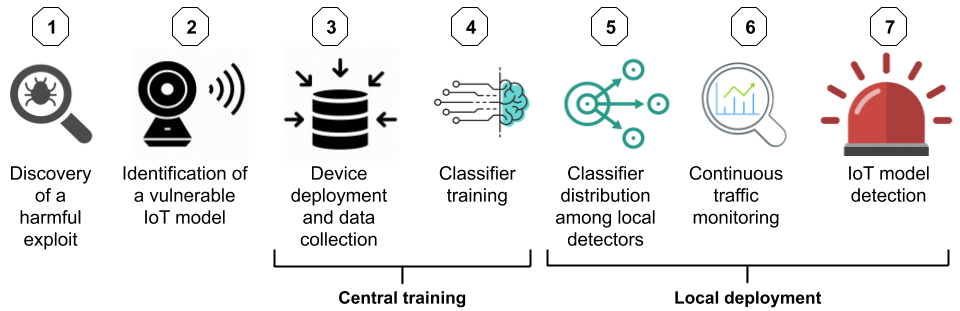}
\caption{An overview of the key steps in our proposed method}
\label{fig:method_overview}
\end{figure}

Whenever a harmful IoT exploit is discovered (step 1 in Figure~\ref{fig:method_overview}), and a related vulnerability is identified (step 2) in a certain IoT model, a telco might want to mitigate the associated risk to its network. To accomplish this, the first step is to detect the presence of such IoT devices among the telco's customers.

In order to detect IoT device models connected behind domestic NATs, we propose a method consisting of \textbf{central training} (steps 3-4) which maximizes efficiency and control, followed by \textbf{local deployment} (steps 5-7) of the trained classifier. We use the following notation for describing our proposed method, which is reproduced for every IoT model separately:

{\small
\begin{compactitem}[leftmargin=*]
    \item[$M$:] Model of an IoT device, defined by the combination of its type, make, and version. 
    For example, \textit{webcam.D\_Link.DCS\_933L} and \textit{webcam.D\_Link.DCS\_\\942L} are two separate IoT models which share the same type and make.
    \item[$D^M$:] Devices $\{ d^M_1, \dots ,d^M_n\}$, instances of IoT model $M$, each defined by a unique MAC address.
    \item[$L$:] Lab 
    which is used for collecting and analyzing traffic data from $D^M$. This central lab can be operated by the telco itself or a third party.
    \item[\(f\):] Flow, an aggregation of the communication between a client and a server, produced using NetFlow, defined by the (1) ingress interface, (2) source IP address, (3) destination IP address, (4) IP protocol, (5) source port, (6) destination port, and (7) IP type of service.
    \item[\(F^M\):] Flow-level dataset, collected in $L$ from $D^M$, using NetFlow and nProbe.
    \item[\(F^M_{training}\), \(F^M_{validation}\):] Training and validation sets for $M$, containing only flows generated by IoT devices of model $M$.
    \item[\(F_{test}\):] Test set, containing flows from various $M$s and also non-IoTs such as PCs, smartphones, etc.
    \item[$C^M$:] One-class classifier for $M$, trained in $L$ using \(F^M_{training}\) and \(F^M_{validation}\).
    \item[$H$:] Homes $\{ h_1, \dots ,h_k\}$, monitored networks of the telco's customers. 
    \item[\(LD\):] Local detectors $\{ ld_1, \dots ,ld_k\}$. Each $ld_i$ is an agent which monitors the NetFlows emerging from the respective home $h_i$ in order to decide whether an IoT device of model $M$ is connected behind $h_i$'s NAT or not.
    \item[\(S_f^{C^M}\):] Anomaly score assigned to a flow $f$ by the one-class classifier $C^M$. The lower this score is, the more chances that $f$ originated from $M$.
    \item[\(th\):] Threshold used to determine if a flow $f$ originated from $M$ or not. 
    Flows with scores below $th$ are classified as originating from non-$M$ devices.
\end{compactitem} 
}


\subsection{Central Training}\label{subsec:central_training}
For a potentially vulnerable IoT model $M$, devices $D^M$ are to be connected to $L$'s internal network behind a NAT. Upon normal usage of $D^M$, network traffic data is generated and subsequently processed by NetFlow, which is installed on the NAT router. 
The produced flows are continuously collected (step 3) using nProbe into a designated server for storage, thus accumulating the raw flow-level dataset $F^M$. Having gathered a sufficient amount of flows in $F^M$, preprocessing steps are to be undertaken, followed by the application of machine learning techniques for training and fine-tuning a classifier $C^M$ (step 4). Instead of conventional binary or multi-class supervised algorithms we propose to train a \textit{one-class} classifier for each $M$ separately. 
With this approach (1) $F^M$ is quickly collected, (2) $C^M$ is trained independently from any non-$M$ device (IoT or not), and (3) $C^M$ can be shared among telcos or other organizations as a standalone classifier. 

\subsection{Local Deployment}\label{subsec:local_deployment}

\begin{figure*}[ht]
\centering
\includegraphics[width=\textwidth, trim={0 12.2cm 0 0},clip]{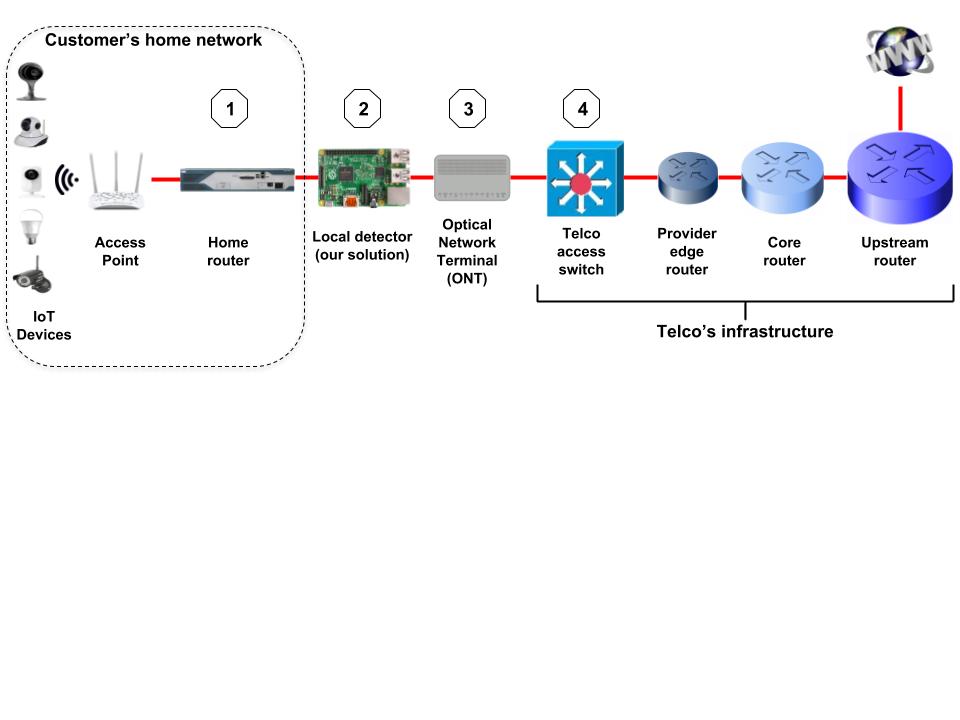}
\caption{Possible locations along the network to deploy the IoT model detection method}
\label{fig:entire_network}
\end{figure*}

In order to preserve the privacy of end users, a telco can deploy a traffic monitoring solution only from \textit{outside} its customers' premises
. This solution cannot be implemented on the home router (location 1 in Figure~\ref{fig:entire_network}) because end users are not obligated to using any type of router. Instead, we propose to place our solution on a hardware agent situated outside the customers' premises, between the home router and the ONT (location 2 in Figure.~\ref{fig:entire_network}). 
This \textit{local detector} monitors the (NATed) traffic data emerging from the home network, and applies the pre-trained 
 classifier in order to detect connected IoT devices of model $M$. 

In step 5 the centrally trained classifier $C^M$ is ditributed among the local detectors $LD$. Each $ld_i$ can be implemented using a low-cost thin computer (such as Raspberry Pi), and should have the following software components: (1) nProbe for collecting flows from $h_i$, (2) a software environment (e.g., Python) for preprocessing the flows, (3) the trained classifier $C^M$ and (4) a software component capable of decision-making and executing actions based on the classification results.
As part of continuous monitoring (step 6) each $f$ collected by $ld_i$ from $h_i$ is preprocessed exactly the way it was in $L$, in order to achieve the same data structure and scale. Then, $f$ is assigned an anomaly score \(S_f^{C^M}\) by $C^M$. 
If \(S_f^{C^M}\) $<$ $th$, $f$ is marked as generated by an IoT device of model $M$ (step 7). 

\subsection{\label{subsec:consequent_actions}Actions to be Triggered by IoT Model Detection}
A positive classification by $ld_i$ can trigger various automated reactions, including:

\textbf{Traffic blocking}: This is probably the most severe reaction, and it is not advised to be taken immediately. The reason is that in false positive detections, traffic blocking means denial of service to legitimate devices, followed by customer dissatisfaction and damage to the reputation of the telco.

\textbf{Email notification}: A more moderate (and perhaps more productive) reaction is to tell the customer that a vulnerable IoT device might be connected to the network, so a software update or a change of password are advised. 
    
\textbf{Additional verification}: Although privacy-preserving, detection that is based only on metadata in a single NetFlow cannot guarantee perfect results (i.e., without any false positives). Therefore, a cascading verification process, in which an additional classifier confirms the detection, can be considered.

We are aware that typically a telco can already see all packet contents including application layer data. Still, we propose the telco to rely solely on the local detectors for data collection, analysis and automated reaction. The main reason is that a central monitoring solution would probably end up with a table that holds information on IoT devices owned by specific customers. On top of the privacy violation, this table could become a valuable goal for attackers.

\section{Evaluation Method}\label{sec:evaluation_ethod}

\subsection{Lab Setup}\label{subsec:lab_setup}
\begin{figure*}[ht]
\centering
\includegraphics[width=0.81\textwidth, trim={0 6.6cm 0 0},clip]{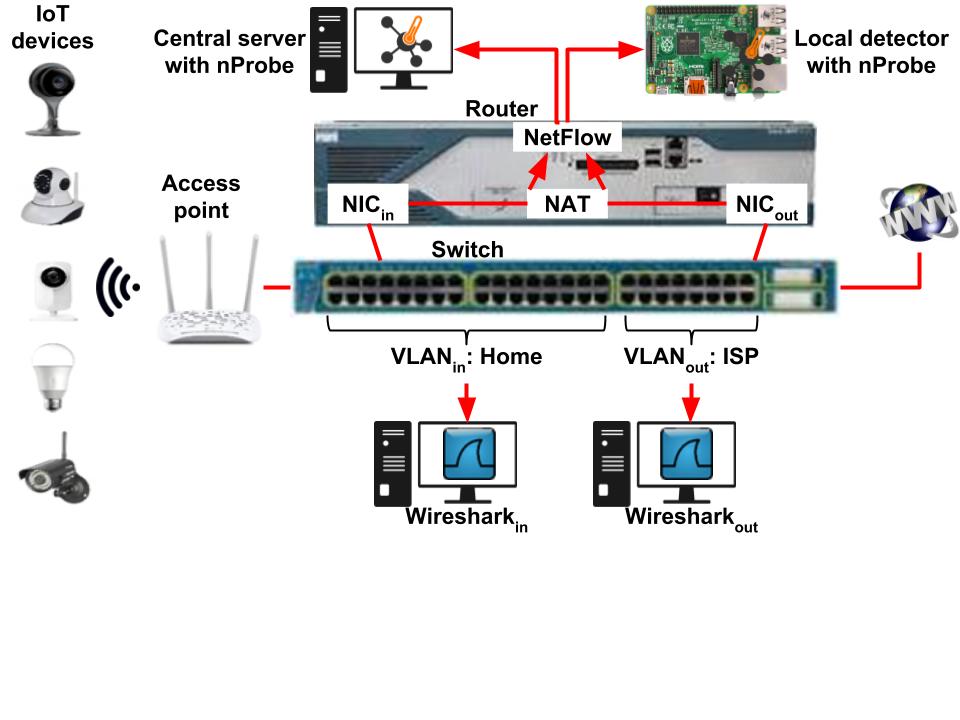}
\caption{Our evaluation setup, imitating a customer's smart home}
\label{fig:experimental_setup}
\end{figure*}

In order to collect representative data, imitating a real-world scenario of various IoT devices connected behind a NAT, we set up a dedicated network as illustrated in Figure~\ref{fig:experimental_setup}.
First, we partitioned a switch into two VLANs: $VLAN_{in}$ and $VLAN_{out}$, representing the home network and the telco side respectively. Then, we connected a variety of commercial IoT devices, as well as laptops and smartphones (details are provided in Table~\ref{tab:data_and_performance}), to $VLAN_{in}$ via a wireless access point. We also connected $VLAN_{in}$ to a NAT router, where NetFlow is already installed. In turn, we connected the router to $VLAN_{out}$, which was connected to the Internet. To imitate the stages of \textit{central training} and \textit{local deployment}, we installed nProbe on a server and a Raspberry Pi respectively, to collect the NetFlow records from the router and analyze them accordingly. In addition, to enable the comparison of our method to previous studies, we also performed port mirroring from $VLAN_{in}$ and $VLAN_{out}$ and captured pcap files using Wireshark.   

\subsection{Data Acquisition}\label{subsec:data_acquisition}
\textbf{Data collection.} We operated the devices routinely over a period of approximately ten days to collect genuine traffic data. For instance, we made the webcams send videos, turned on and off the sockets and the light bulb, surfed the Web via the laptops and smartphones, etc. The resultant traffic was captured simultaneously using NetFlow and Wireshark.

\textbf{Ground truth labeling.} 
In our lab we recorded flows both behind and in front of the NAT, and we also made sure to configure static IPs in the internal network. This way we were able to match each external outbound flow (labeled with the source IP address of the router) with its internal twin, correctly labeled with the source IP. By matching these with a table of IP/MAC/device model, we labeled the NetFlows for analysis. We repeated the same matching procedure with the pcap files, so the ground truth labels are available for them as well.

\textbf{Feature extraction.}
We used the following NetFlow features~\cite{netflow2011format} as a feature set that is minimal yet potentially informative for IoT model detection:
\begin{enumerate}
    \item IN\_BYTES: The number of incoming bytes associated with an IP Flow
    \item OUT\_BYTES: The number of outgoing bytes associated with an IP Flow
    \item DST\_TOS: Type of Service byte setting when exiting outgoing interface
    \item SRC\_TOS: Type of Service byte setting when entering incoming interface
    \item PROTOCOL: IP protocol byte
    \item L4\_DST\_PORT: TCP/UDP destination port number
    \item L7\_PROTO\_NAME: Layer 7 protocol name
    \item flow\_duration: Extracted from NetFlow by subtracting FLOW\_START\_ \\ MILLISECONDS from FLOW\_END\_MILLISECONDS
\end{enumerate}

\subsection{Preprocessing and Experimentation}\label{subsec:preprocessing_experimentation}
First, we partitioned $F$ chronologically, such that the earliest 70\% of each device (identified by the MAC address) is included in $F^M_{training}$, the next 10\% in $F^M_{validation}$ and the remaining (latest) 20\% in $F_{test}$. Then, using Python and Scikit-Learn, we repeated the following experiment 13 times, once for each $M$:
\begin{enumerate}
    \item We filtered out all of the non-$M$ flows from both $F^M_{training}$ and $F^M_{validation}$, because we chose the technique of one-class classification.
    \item In $F^M_{training}$ we scaled the numeric features to the range of [0,1] and encoded the categorical features into dummy variables of the same range.
    \item While preprocessing $F^M_{validation}$ and $F_{test}$ we performed the same scaling and encoding, specific to the current $M$. Consequently, the number of dummy variables differs among IoT models, depending on the number of unique values found in $F^M_{training}$ for each categorical feature (e.g., L4\_DST\_PORT).
    \item We trained $C^M$ on $F^M_{training}$ and saved it to the disk to enable distribution to the local detectors. In preliminary experimentation we found that the Isolation Forest algorithm performs much better than One-Class SVM and Local Outlier Factor (LoF), so we chose it as the sole algorithm in our study.
    \item We applied $C^M$ to $F_{test}$ to evaluate the method's classification performance.
\end{enumerate}

\subsection{Performance Metrics}\label{subsec:performance_metrics}
We used the following widely-accepted metrics to evaluate our detection method:
\begin{enumerate}
    \item True Positive Rate (TPR): Ratio of cases where $f$ was generated by $M$ and truthfully detected as such by $C^M$. 
    \item False Positive Rate (FPR): Ratio of cases where $f$ was not generated by $M$ but falsely classified as such by $C^M$.
    \item The area under the ROC curve (ROC AUC): Class discrimination capability for differing $th$ values.
    \item Time to detect: Depends on the interarrival time (IAT) of $f^M$, as well as its duration, preprocessing, and classification times.  
\end{enumerate}

\section{Results and Discussion}\label{sec:Results_and_discussion}

\begin{table}
  \caption{The IoT models in our experiments, their datasets and the classification performance using $F_{test}$}
  \label{tab:data_and_performance}
  \includegraphics[width=\linewidth]{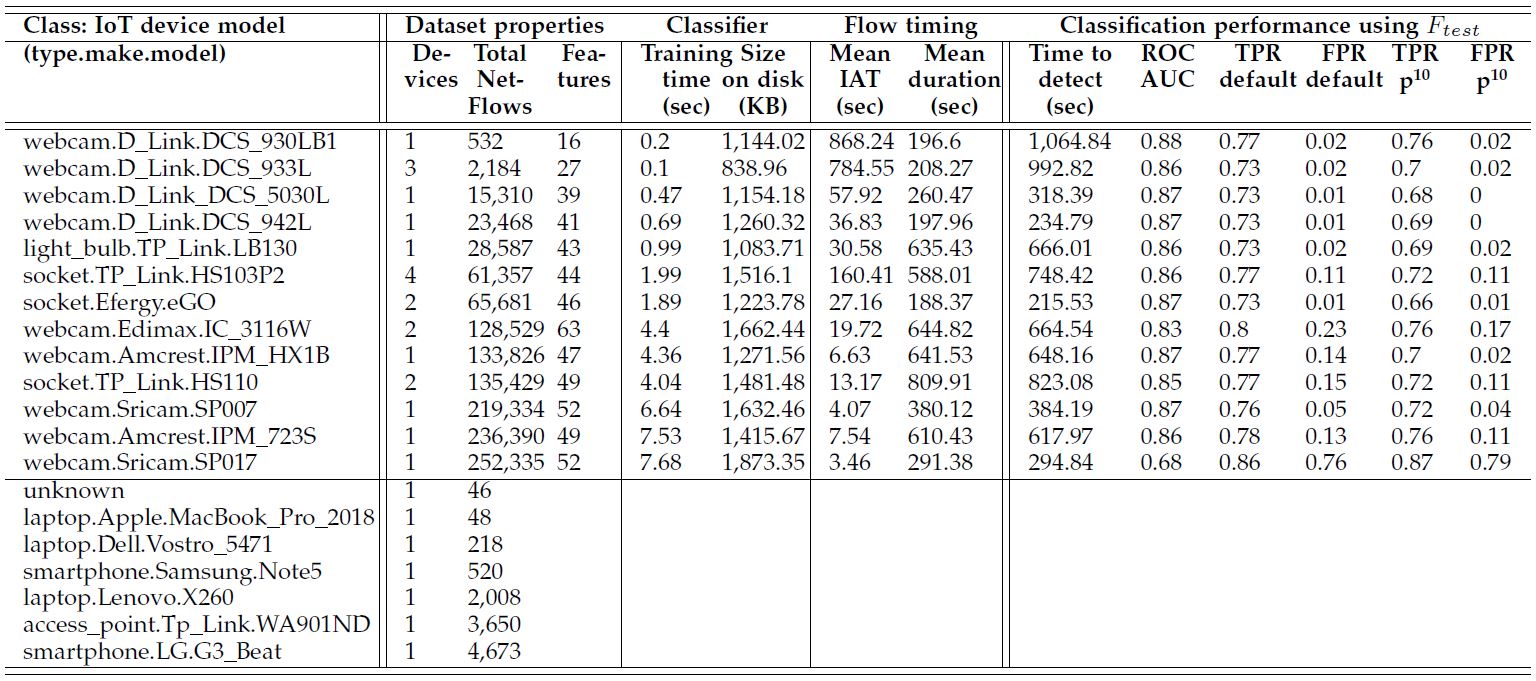}
\end{table}

\subsection{Experimental Results}\label{subsec:experimental_results}
\begin{figure}\small
\begin{minipage}[b]{.55\linewidth}
\centering
\includegraphics[width=\textwidth, valign=t, trim={0 0 0 1cm},clip]{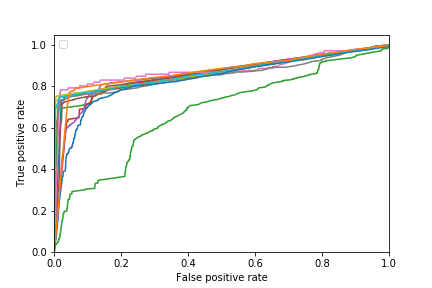} 
\subcaption{\strut (a) ROC curves for multiple IoT \\ models $M$ using $F_{test}$, based on classifying \\ NetFlows}
\label{fig:roc_curves_single}
\end{minipage}%
\begin{minipage}[b]{.55\linewidth}
\centering
\includegraphics[width=\textwidth, valign=t, trim={0 0 0 0.85cm},clip, scale=0.95]{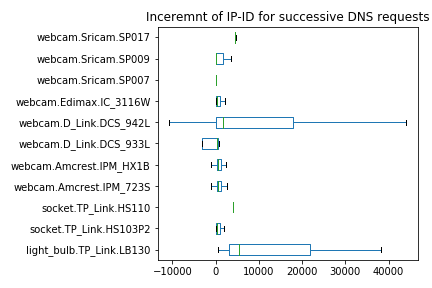}

\subcaption{\strut (b) Distribution of the increment in IP-ID values of DNS requests for IoT models in our experiment}\label{fig:ip_id_increment_distribution}

\end{minipage}

\vskip\baselineskip
\begin{minipage}[b]{.55\linewidth}
\centering
\includegraphics[width=\textwidth, valign=t, scale=0.75]{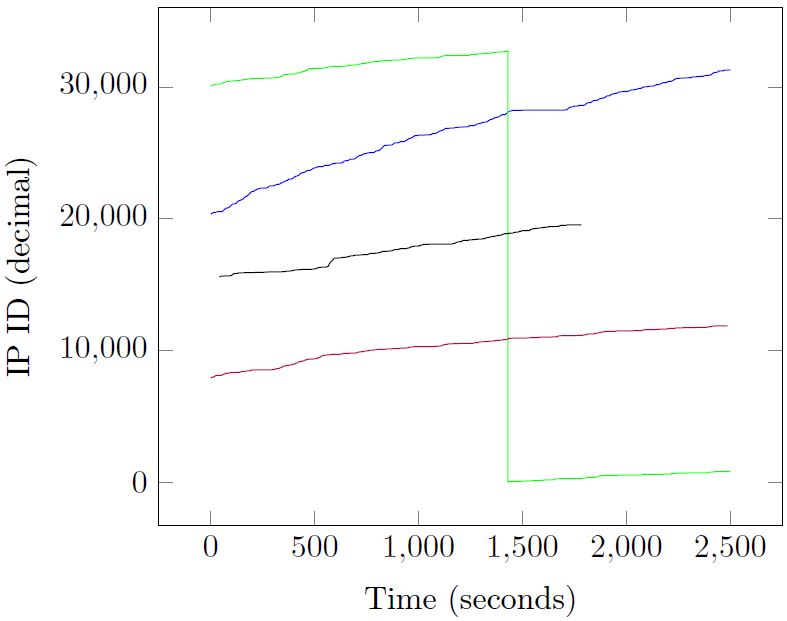} 

\subcaption{\strut (c) DNS IP-ID of four Windows- \\ and Android-based hosts over time \\ (adapted from~\cite{orevi2018dns})}
\label{fig:dns_ip_id_orevi}%
\end{minipage}
\begin{minipage}[b]{.55\linewidth}
\centering
\includegraphics[width=\textwidth, trim={0 0 0 0.85cm},clip, valign=t]{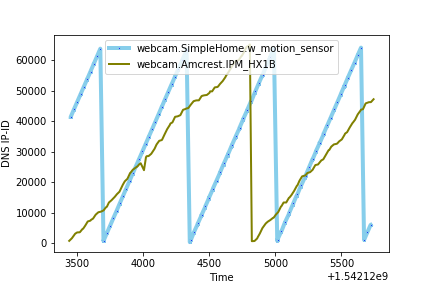}
\subcaption{\strut (d) DNS IP-ID of two Linux-based IoT models over time (gathered from our experiments), motivating the "slope-matching" idea}\label{fig:dns_ip_id_ours}
\end{minipage}
\caption{Experimental results using $F_{test}$
}\label{fig:results_figures}
\end{figure}

Our experiments are summarized in Table~\ref{tab:data_and_performance}, sorted by the number of NetFlows. Excluding the bottom seven rows (non-IoTs), leads to the following conclusions:
\begin{itemize}
    \item \textbf{Training time.} The mean (standard deviation) of the time it takes to train $C^M$ is only 3.15 (2.81) seconds. Thus, frequent retraining for performance improvements is highly feasible. 
    \item \textbf{Classifier size.} $C^M$ requires very little disk space - just 1,350.62 (280.39) KB. Thus, deploying it on thin local detectors is practical.
    \item \textbf{Time to detect.} The time it takes to preprocess a given $f$ is in the order of microseconds, and so is the classification time; thus, both are negligible. The IAT and duration of $f$ are much more significant, and their sum (shown in Table~\ref{tab:data_and_performance}) varies between approximately 4 and 18 minutes.  
    \item \textbf{ROC AUC.} For most $M$s, reasonable values of 0.85 (0.05) are attained using $F_{test}$. Only \textit{webcam.Sricam.SP017} performs substantially worse; a closer look revealed that in about 19\% of cases it is confused with \textit{webcam.Amcrest.IPM\_\\723S}. Apparently, the reason for this confusion is the substantial overlap in their communicated domains, including Amazon, HTTP.Amazon, NTP, NTP.Amazon, and SSL.Amazon. Figure~\ref{fig:results_figures}(a) shows how almost all of the ROC curves (one for each $M$) share the same shape.
    
    \item \textbf{Default TPR and FPR. }
     When $f$'s classification is determined based on comparing $S_f^{C^M}$ to Scikit-Learn's default $th$, a TPR of 0.76 (0.04) is obtained. 
     This means that, on average, when an IoT device of model $M$ is connected behind a NAT, a telco can detect it in 76\% of the cases based only on metadata captured in a single NetFlow.
     Consequent actions (see Subsection~\ref{subsec:consequent_actions}) can then substantially mitigate the risk to the telco's infrastructure and service. 
     Unfortunately, this TPR is accompanied by an FPR of 0.13 (0.20), meaning that in too many cases false alarms are generated. Hence, we looked for solutions to reduce the FPR while preserving satisfactory TPR levels. 
    \item $\mathbf{S_f^{C^M}}$\textbf{-percentile-based TPR and FPR.} While searching for methods to overcome the challenge of high FPR, we noticed that using $F^M_{validation}$ for $th$ calibration can improve the classification performance using $F_{test}$. That is, we (1) trained the classifier $C^M$ using $F^M_{training}$, (2) applied $C^M$ to $F^M_{validation}$, (3) calculated a predefined percentile of the resultant distribution of the score $S_f^{C^M}$ and (4) used this value as the new $th$ to act as a classification threshold when using $F_{test}$. 
    So, for varying percentiles [0, 30] of $S_f^{C^M}$ on $F^M_{validation}$ we recalculated the TPR and the FPR using $F_{test}$, and we found that the $10^{th}$ percentile (annotated $P^{10}$ in Table~\ref{tab:data_and_performance}) provides the best results: A decrease of FPR for almost all the IoT models, and most substantially for \textit{webcam.Edimax.IC\_3116W}, \textit{webcam.Amcrest.IPM\_HX1B} and \textit{socket.TP\_Link.HS110}. Actually, for two $M$s the FPR decreased to absolute zero, and for five others the FPR was found to be 0.02 or less. Overall, the FPR decreased to 0.11 (0.21) with a cost of reducing the TPR to 0.73 (0.05). We note that this level of performance is yet to be improved in order to support actual deployment by a telco, and we discuss it in Subsection~\ref{subsec:limitations}. 
\end{itemize}

\subsection{Benchmarking}\label{subsec:benchmarking}
In order to evaluate our proposed method more comprehensively, we decided to empirically compare it with two current deNATing methods. We implemented them in our lab with a few necessary adjustments, and because they use packet-level traffic data we tested them on pcap files that had been collected simultaneously by Wireshark.

\subsubsection{DNS IP-ID-Based deNATing}\label{subsubsec:dns_based}

The method proposed in~\cite{orevi2018dns} aimed at deNATing traffic of devices using the same OS, rather than detecting different IoT models. It relies on the fact that the IP-ID field in some OSs is consistently incremented for successive packets sent to the same destination IP. Since DNS requests to the same DNS resolver are sent to the same destination IP, the value of the IP-ID field is monotonically increasing between DNS requests. Thus, tracking IP-ID values of multiple DNS requests coming out of an NAT router to the same resolver may assist in correlating subsequent packets of distinct devices behind the NAT.

The researchers experimented with Windows 8 and 10, and Android. They showed that for successive DNS requests the difference in IP-ID values is very stable (see Fig.~\ref{fig:results_figures}(c)) and almost always equals one. This makes deNATing rather simple and effective with these OSs, unlike our Linux-based IoTs, where the difference was found to be highly variable (see Fig.~\ref{fig:results_figures}(b)). Still, as can be seen in Fig.~\ref{fig:results_figures}(d), different IoT models may increment their respective DNS IP-ID values in a consistent and characteristic way, such that robust linear regression models can be trained. In turn, the trained slopes can be compared with the ones found in a test set. A good match between trained and observed slopes can be the basis for an IoT model detection technique. We leave this to future work.

\subsubsection{DNS Domain-Based deNATing}\label{subsubsec:dns_domain_based}
In~\cite{guo2018ip} the objective was to identify the type of NATed IoT devices, similarly to this paper. For each IoT device the authors tracked a list of communicated ~\textit{device-facing} server names. Then, during a test period, if the number of communicated device server names from the related list surpassed a threshold, they inferred that the device type is present. Their method is somewhat limited as it poses the following constraints on the type and quantity of the communicated servers: 

\begin{itemize}
    \item It excludes \emph{third party} and \emph{human-facing} server names to minimize the FPR.
    \item It excludes device types which communicate with less than three servers.
    \item If an overlap exists among the server names of multiple device types, the method cannot guarantee that the devices are distinguishable, and it reports that at least one of them was detected.
\end{itemize}

\begin{figure}[ht]
\centering
\includegraphics[width=0.7\textwidth]{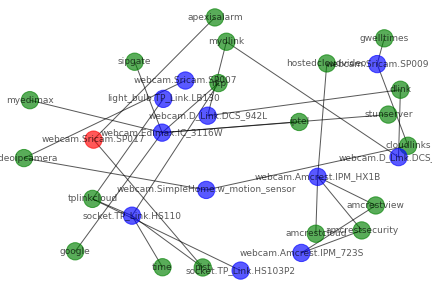}
\caption{The domains (green nodes) requested by our IoT devices (blue). The red node represents \textit{webcam.Sricam.SP017}, which had the lowest ROC AUC.}
\label{fig:network_chart_devices_domains}
\end{figure}
 
Fig.~\ref{fig:network_chart_devices_domains} illustrates the results of applying the method proposed in~\cite{guo2018ip} to our test set. It presents the DNS domains (green nodes) requested by the IoT devices (blue nodes) in our experiment, where it can be seen that some of them are human-facing, some belong to third parties, and also an overlap of requested domains clearly exists in our experiment. In practice, meeting all of the above constraints would eliminate all the devices we experimented with, so we implemented the DNS domain-based method without limiting the types of communicated servers. We also used the server names rather than their resolved IP addresses, in order to promote efficiency, while relying on the fact that Linux-based IoT devices don't support DNS caching. Also, in the original paper they reported the detection performance on a time window of days. To compare the performance of their method to ours more fairly, we examined it using a time window of 10 minutes. 

Altogether, only three of our IoT models met the criterion of at least three device-facing communicated servers, and they performed well: a TPR of 1.00 for all of them, an FPR of 0.00 for \textit{socket.TP\_Link.HS110} and \textit{webcam.Amcrest.IPM\_\\HX1B}, and an FPR of 0.56 for \textit{webcam.Edimax.IC\_3116W}. However, a coverage of three out of 13 IoT models seems far from sufficient for a telco to implement.

\subsection{\label{subsec:limitations}Limitations}

We are aware of some shortcomings of our proposed method, as follows. First, upon deployment, a telco can train $C^M$ only after it purchases devices of model $M$, configures them and collects a sufficient amount of  NetFlow records.
This process might take few days to complete.
However, we are not aware of any other traffic-based detection method that can skip the time-consuming data acquisition stage. 
To shorten this stage it is advised to connect multiple $M$ devices.
Second, firmware updates to IoT models might make the classifiers obsolete.
Again, any other data-driven classifier would probably face the same challenge.

\subsection{\label{subsec:future_research}Future Research}
The scope of this paper was limited to developing a method which is capable of \textit{detecting} vulnerable IoT models behind a NAT. However, detection is only a first stage, to be complemented with locally-installed tools such as \textit{vulnerability scanning} (to check if the vulnerability has been patched) or \textit{virtual patching}.

Additional challenges to address are (1) improving our method in terms of the TPR and FPR, possibly using the cascading detection approach (discussed in Subsection~\ref{subsec:consequent_actions}), (2) looking for solutions that are less costly than local detectors, can support a multitude of households and are still privacy preserving, and (3) exploring the potential of the DNS IP-ID "slope matching" idea.

\section{Conclusion}
\label{sec:conclusions}

In this paper, we demonstrated that using our proposed method enables a telco to detect about 73\% of any NATed IoT model of interest. This is a first step towards dramatically mitigating the risk posed to the telco's infrastructure by domestic IoT devices that might be recruited to botnets. The detection takes only few minutes, and is being performed while preserving customers' privacy.



\bibliographystyle{splncs04}
\bibliography{Mendeley_ESORICS_references.bib}

\end{document}